\documentclass[prl,preprint,doublecolumn,superscriptaddress]{revtex4-1}  

\usepackage{graphicx}
\usepackage{verbatim}
\usepackage{float}
\usepackage{sidecap}
\usepackage{dcolumn}
\usepackage{bm}
\usepackage{amssymb}
\usepackage{amsmath}	
\usepackage{color}
\usepackage{multirow}
\usepackage{lipsum}
\usepackage[a4paper]{geometry}
\newgeometry{top=2cm,right=1.25cm,left=1.25cm,bottom=1.8cm}

\draft
\begin{document}
\title{Strain and Electric Field Control of Hyperfine Interactions for Donor Spin Qubits in Silicon}

\author{M. Usman} 
\email{musman@unimelb.edu.au} 
\affiliation{Center for Quantum Computation and Communication Technology, School of Physics, The University of Melbourne, Parkville, 3010, VIC, Australia.}

\author{C.D. Hill} 
\affiliation{Center for Quantum Computation and Communication Technology, School of Physics, The University of Melbourne, Parkville, 3010, VIC, Australia.}

\author{R. Rahman} 
\affiliation{Electrical and Computer Engineering Department, Purdue University, West Lafayette, IN, USA.}

\author{G. Klimeck} 
\affiliation{Electrical and Computer Engineering Department, Purdue University, West Lafayette, IN, USA.}

\author{M.Y. Simmons} 
\affiliation{Center for Quantum Computation and Communication Technology, School of Physics, The University of New South Wales, Sydney, 2052, NSW, Australia.} 

\author{S. Rogge} 
\affiliation{Center for Quantum Computation and Communication Technology, School of Physics, The University of New South Wales, Sydney, 2052, NSW, Australia.}  

\author{L.C.L. Hollenberg} 
\affiliation{Center for Quantum Computation and Communication Technology, School of Physics, The University of Melbourne, Parkville, 3010, VIC, Australia.} 

\begin{abstract}
Control of hyperfine interactions is a fundamental requirement for quantum computing architecture schemes based on shallow donors in silicon. However, at present, there is lacking an atomistic approach including critical effects of central-cell corrections and non-static screening of the donor potential capable of describing the hyperfine interaction in the presence of both strain and electric fields in realistically sized devices. We establish and apply a theoretical framework, based on atomistic tight-binding theory, to quantitatively determine the strain and electric field dependent hyperfine couplings of donors. Our method is scalable to millions of atoms, and yet captures the strain effects with an accuracy level of DFT method. Excellent agreement with the available experimental data sets allow reliable investigation of the design space of multi-qubit architectures, based on both strain-only as well as hybrid (strain+field) control of qubits. The benefits of strain are uncovered by demonstrating that a hybrid control of qubits based on (001) compressive strain and in-plane (100 or 010) fields results in higher gate fidelities and/or faster gate operations, for all of the four donor species considered (P, As, Sb, and Bi). The comparison between different donor species in strained environments further highlights the trends of hyperfine shifts, providing predictions where no experimental data exists. Whilst faster gate operations are realisable with in-plane fields for P, As, and Sb donors, only for the Bi donor, our calculations predict faster gate response in the presence of both in-plane and out-of-plane fields, truly benefiting from the proposed planar field control mechanism of the hyperfine interactions.
\end{abstract}


\maketitle

\section{1. Introduction}

During the last few years, there has been significant progress~\cite{Fuechsle_NN_2012, Weber_Nature_Nano_2014, Pla_Nature_2013, Saeedi_Science_2013, Muhonen_NatureNano_2014} towards the realization of quantum computing architectures based on shallow donors in silicon (Si)~\cite{Kane_Nature_1998, Hollenberg_PRB_2006}. Several techniques have been explored to implement precise control of the nuclear or electronic spins through wave function engineering of donors using either electric fields~\cite{Lansbergen_Nat_Phys_2008} or strain fields~\cite{Huebl_PRB_2006, Dreher_PRL_2011}. At the core of such approaches, controlled manipulation of the donor hyperfine coupling is a critical component. Previous theoretical studies have been primarily focused on the electric-field-dependent Stark shift of the hyperfine interaction~\cite{Kane_Nature_1998} for donors in Si, which is now well understood from both theory~\cite{Martins_PRB_2004, Rahman_PRL_2007, Usman_JPCM, Pica_PRB_2014, Zwanenburg_RMP_2013} and experiments~\cite{Bradbury_PRL_2006, Lo_arxiv_2014, Pica_PRB_2014}. In comparison, the strain-dependence of the hyperfine coupling is relatively less studied, despite offering a promising alternative to manipulate the hyperfine coupling of donors. The presence of strain, in contrast to the use of an electric field, eliminates the possibility of ionization, as the control of the donor wave function is mechanical rather than electrostatic. Additionally strain can drastically reduce valley oscillations of exchange coupling~\cite{Koiller_PRB_2002, Wellard_Hollenberg_PRB_2005}, which would play an important role in field control of qubits in strained environments. Recent progress towards atomically precise fabrication of donors in strained Si provides a testbed to demonstrate the advantages of strain in the realization of donor-based qubit devices~\cite{Lee_Nanotech_2014}. Whilst the previous studies have exclusively considered strain or electric field effects on the quantum control of the donors, it is clear that through valley physics there is a subtle interplay between these two effects. This work establishes a multi-scale theoretical approach to provide an understanding of the impact of strain and electric fields simultaneously present in the qubit devices, and predicts that such a hybrid quantum control scheme can open new avenues for architectures with faster single spin gates and spin-dependent tunneling read-out strategies.       

Existing theoretical studies of the impact of strain on the hyperfine interaction of donors have been based on either valley-repopulation model (VRM) derived from effective-mass theory (EMT)~\cite{Wilson_PR_1961, Koiller_PRB_2002, Fritzche_PR_1962, Hale_PR} or density functional theory (DFT)~\cite{Huebl_PRB_2006}. While the VRM model was useful in providing a first-order description of the hyperfine shifts for small strain fields, it failed to explain the experimentally measured hyperfine reduction at large strain fields~\cite{Huebl_PRB_2006}. The DFT calculations for strained Si:P exhibited good agreement with the experimental measurements for an extended range of strain fields~\cite{Huebl_PRB_2006}, highlighting the importance of atomistic approaches. However, this method is limited to only few-atom systems, and is consequently unable to reproduce the donor binding energy spectra and provide a detailed picture of the wave functions~\cite{Overhof_PRL_2004}. Therefore the requirement for a theoretical framework with an atomistic accuracy accompanied by scalability to large-scale realistic systems remains a critical challenging problem. 

Our work fills this theory gap by establishing a multi-scale atomistic tight-binding framework, which in contrast to other approaches~\cite{Koiller_PRB_2002, Klimeck_1, Klimeck_2, Laucht_arxiv_2015} explicitly includes central-cell corrections and non-static dielectric screening of the donor potential. The multi-million-atom simulations for strain-dependent hyperfine interaction are benchmarked to a high level against both ab-initio approaches and experiment. The electric field dependence of the hyperfine is accurately captured by demonstrating excellent agreement with the experimentally measured Stark shift data~\cite{Pica_PRB_2014} for all of the four donor species considered (P, As, Sb, and Bi).  A clear understanding of the influence of strain on the physical properties of the donor is presented in terms of underlying valley physics. The performance prospects of unstrained and strained Si substrates are explored, uncovering the benefits of strain for qubit devices, in particular by showing that a hybrid control of qubits based on (001) compressive strain and in-plane fields (100 or 010) results in higher gate fidelities and/or faster gate operations for all of the four donor species. Due to recent research interests for As, Sb, and Bi donors~\cite{Bradbury_PRL_2006, Wolfowicz_NatureN_2013, Lo_arxiv_2014}, we also present a comparison among different donor species in strained Si environments, further highlighting the trends of strain and electric field induced shifts in the hyperfine couplings, and providing predictions at large strain fields where no previous experimental or theoretical data exists. A novel scheme of two-dimensional hyperfine control in strained environments is explored based on electric fields from top and side gates. Whilst faster gate operations are realisable with the in-plane fields for P, As, and Sb donors, only for the Bi donor, faster gate response is predicted in the presence of both in-plane (100 or 010) and out-of-plane (001) fields, truly benefiting from the proposed planar field control mechanism of hyperfine control.

\section{2. Theoretical Framework}

In our TB approach, the Si bulk band structure is reproduced using a twenty-orbital (sp$^3$d$^5$s$^*$) basis~\cite{Boykin_PRB_2004}. The donor atom is placed at the center of a large Si box (40$\times$40$\times$40 nm$^3$) consisting of roughly 3.1 million atoms, and is represented by a Coulomb potential, U(r), which is screened by non-static dielectric function for Si and is given by:

\begin{equation}
	\label{eq:Nonstatic_donor_potential}
	U \left( r \right) = \frac{-e^2}{ \epsilon  r} \left( 1 + A \epsilon \mathrm{e}^{- \alpha r} + \left( 1-A \right) \epsilon \mathrm{e}^{- \beta r} - \mathrm{e}^{- \gamma r}  \right)
\end{equation}

\noindent
\\ where $e$ is the electronic charge, and the previously published values of $\epsilon$, $A$, $\alpha$, $\beta$, and $\gamma$ are taken from the literature~\cite{Usman_JPCM, Note_1}. Recently we have demonstrated the importance of the central-cell corrections and the non-static dielectric screening of the donor potential to accurately reproduce the experimental Stark shift for the Si:As donors~\cite{Usman_JPCM}. We now show that the non-static dielectric screening is also crucial to accurately reproduce the experimentally measured strained hyperfine interaction at large strain fields. Therefore this work extends the TB model using the non-static screening function to P, Sb, and Bi donors. The donor potential is truncated to U$_0$ at the donor site, $r$=0. The values of U$_0$ are adjusted to reproduce the experimentally measured binding energy spectra of the donors~\cite{Ramdas_RPP_1981}. By using the U$_0$ values of 3.5 eV, 2.2 eV, 3.8365 eV, and 4.6668 eV for P, As, Sb, and Bi donors respectively, we calculate the binding energies of the ground states A$_1$ within 1 $\mu$eV and the binding energies of the excited states (T$_2$ and E) within 1 meV of the experimental values for all of the four donor species. It should be emphasized that whereas multi-valley EMT theories have been successful in fitting the ground state binding energies~\cite{Pica_PRB_2014}, the fitting of all of the three 1s states simultaneously with such a level of accuracy has been inaccessible. The accurate fitting of the excited state energies, as achieved in this work, is critically important for strain-dependent hyperfine studies, where the excited state E mixes with the ground state A$_1$ as a function of the strain.

\begin{figure*}[t!]
\centering
\includegraphics[height=12cm, width=14.5cm]{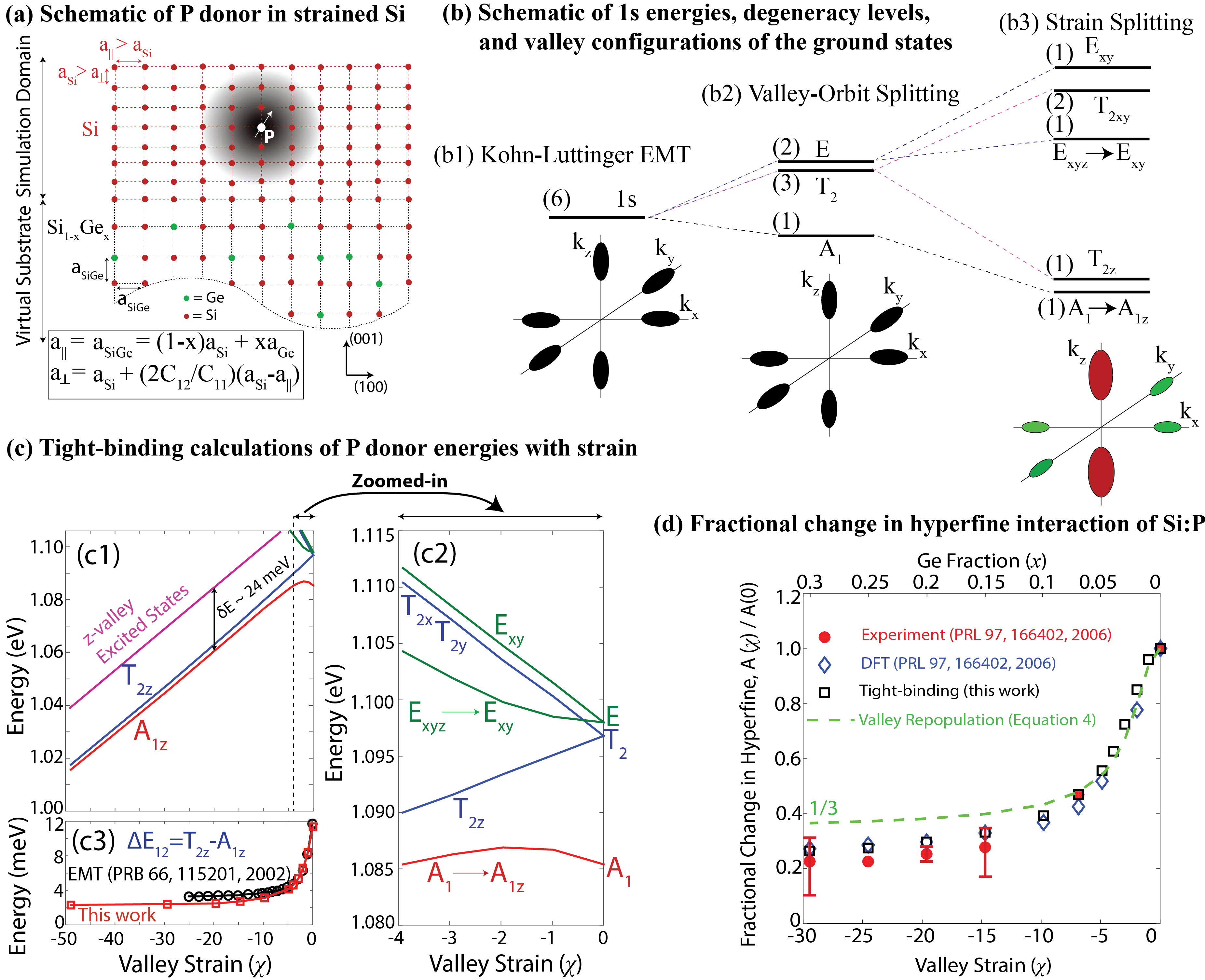}
\caption{\textbf{Benchmarking tight-binding theory against experiment and DFT method for P donor:} (a) An artist's view of a single P donor in the strained Si. (b) Schematic diagram to indicate the impact of (b2) valley-orbital interaction and (b3) strain on the splitting of 1s energy levels and the valley configuration of the lowest ground state A$_1$. The degeneracy of each energy level is also labeled.  (c) The energies of the lowest few states of the P donor in the strained Si as a function of the valley strain ($\chi$), with (c1) for a large variation of strain and (c2) the low strain region ($-$4 $< \chi <$ 0) is zoomed-in to highlight the valley repopulation regime. (c3) The energy difference of the lowest two states as a function of $\chi$. (d) The computed fractional change in the hyperfine interaction A($\chi$)/A(0) is plotted as a function of $\chi$, and is compared with the previously published experimental measurements, and calculations based on valley-repopulation model and DFT.}
\label{fig:Fig1}
\end{figure*}  

\section{3. Results and Discussion}

For the study of the strain-dependent hyperfine interaction of donors, we first benchmark our model against the recent experimental data set for the P donors~\cite{Huebl_PRB_2006}. A commonly adopted procedure to induce strain is by using the lattice mismatch technique, where two materials with different lattice constants are grown on top of each other. Such a technique is depicted in Figure~\ref{fig:Fig1}(a), where P doped Si is shown on top of a Si$_{1-x}$Ge$_x$ virtual substrate. The amount of strain in the Si region can be tuned by varying the Ge fraction ($x$) in the substrate. The lattice constant of Si$_{1-x}$Ge$_x$ is larger than Si, and therefore induces a tensile strain in P-doped Si region along the in-plane directions (a$_{||} >$ a$\rm _{Si}$). Consequently, the out-of-plane lattice constant (a$_{\perp}$) shrinks in accordance with the Poisson's ratio, leading to a compressive strain along the growth direction (a$_{\perp} <$ a$\rm _{Si}$). For a (001)-oriented Si$_{1-x}$Ge$_x$/Si system, the growth axis is the z-axis and the growth plane is the xy-plane, implying that the z-valleys (xy valleys) will primarily experience the effect of a compressive (tensile) strain.
\\ \noindent
\textbf{3.1 Characterising strain effects through valley physics:} \\ Figure~\ref{fig:Fig1}(b) schematically illustrates different effects on the splittings of 1s donor states. The valley configuration of the lowest energy ground state is also included. A simple effective mass theory without multi-valley effects, such as presented by Kohn-Luttinger~\cite{Kohn_PR_1955}, would lead to a six-fold degenerate 1s state as shown in Figure~\ref{fig:Fig1}(b1). In reality, the effect of valley-orbit interactions results in a splitting of the 1s energies into three sets (Figure~\ref{fig:Fig1}(b2)). The lowest ground state A$_1$ is a singlet state, which is made up of all six valleys with a configuration of $\frac{1}{\sqrt{6}}$\{1, 1, 1, 1, 1, 1\}. The first triply-degenerate excited state (T$_2$) has the following valley configurations: T$_{2x}$=$\frac{1}{\sqrt{2}}$\{1, -1, 0, 0, 0, 0\}, T$_{2y}$=$\frac{1}{\sqrt{2}}$\{0, 0, 1, -1, 0, 0\}, and T$_{2z}$=$\frac{1}{\sqrt{2}}$\{0, 0, 0, 0, 1, -1\}. The second doubly-degenerate excited states (E) are composed of the following valley configurations: E$_{xy}$=$\frac{1}{2}$\{1, 1, -1, -1, 0, 0\} and E$_{xyz}$=$\frac{1}{\sqrt{12}}$\{-1, -1, -1, -1, 2, 2\}. The influence of strain on the donor energies can be understood in terms of their valley configurations: the valleys in the direction of compressive strain experience a reduction in energy (higher population) and that valleys in the direction of tensile strain exhibit an increased energy (lower population)~\cite{Wilson_PR_1961}. Since the excited states consist of assymetric valley contributions, they experience different effects of strain and therefore do not remain degenerate in the presence of strain (Figure~\ref{fig:Fig1}(b3)). In our case, the tensile strain along the x and y directions will push the states with xy-valley configurations (T$_{2x}$, T$_{2y}$, E$_{xy}$) up and the compressive strain along the z-axis will shift the states with z-valley configurations (A$_{1z}$ and T$_{2z}$) downward on the energy scale. The valley repopluation effect for A$_{1z}$ has also been shown by illustrated by showing z-valleys (indicated by the red color) larger in size when compared with the xy valleys (indicated by the green color). in the schemetic diagram of Figure~\ref{fig:Fig1}(b3).   

The impact of strain on the donor states can be characterized either directly in terms of the Ge fraction $x$ in the substrate, or can be described in terms of a dimensionless parameter so called the valley strain, $\chi$, which was derived by Wilson and Feher~\cite{Wilson_PR_1961}, and is given by: 

\begin{equation}
	\label{eq:valley_strain}
	 \chi = \frac{\Xi_u}{3 \Delta_c} \left( \frac{a_{\rm Si} - a_{\rm Ge}}{a_{\rm Si}} \right) \left( 1 + \frac{2 \rm C_{12}}{\rm C_{11}} \right) x  
\end{equation}

\noindent 
Here the value of the uniaxial strain parameter $\Xi_u$ is 8.6 eV, C$_{11}$ and C$_{12}$ are the elastic constants of Si and the value of their ratio C$_{12}$/C$_{11}$ is 2.6, 6$\Delta_c$=12.96 eV is the energy splitting of the singlet (A$_1$) and doublet (E) states for the unstrained bulk P donor, a$_{\rm Si}$=0.5431 nm and a$_{\rm Ge}$=0.5658 nm are the bulk Si and Ge lattice constants respectively, and $x$ is the concentration of Ge in the virtual Si$_{1-x}$Ge$_x$ substrate. For this study, we vary $x$ between 0 and 0.5, which corresponds to a variation of $\chi$ from 0 to $\approx-$49. For each value of $x$, a strained TB Hamiltonian~\cite{Boykin_PRB_2002} is solved to compute the strained donor energies and states.   

Figure~\ref{fig:Fig1}(c1) plots the P donor energies calculated from TB simulations for a large variation of the valley strain $\chi$. The valley repopulation effect primarily occurs for small magnitudes of $\chi$, so Figure~\ref{fig:Fig1}(c2) presents zoomed-in version of the plot for $-$4 $< \chi <$ 0 to highlight this effect. The effect of applied strain is on donor energies is accurately captured by the TB theory, indicating a partial lift of the degeneracy of the T$_2$ states, splitting them into a single T$_{2z}$ state whose energy decreases due to compressive strain along the z-axis, and a pair of degenerate T$_{2x}$ and T$_{2y}$ states with their energies increasing due to the effect of tensile strain. The T$_{2x}$ and T$_{2y}$ states remain degenerate as the same magnitude of strain is applied along the x and y-axis (a$_x$=a$_y$=a$_{||}$). The strain completely lifts the degeneracy of the E states. The energy of the E$_{xy}$ state increases due to the effect of tensile strain.   

The remaining two states A$_1$ and E$_{xyz}$ with contributions from all of the six valleys experience the effect of both compressive strain along the z-axis and the tensile strain along the x and y-axis, and therefore exhibit a nonlinear dependence on $\chi$. The strain mixes the E$_{xyz}$ excited state into the ground A$_1$ state. For $-$4 $< \chi <$ 0, the A$_1$ state experiences the competing effects of the tensile and compressive strains. Initially the increase in $a_{||}$ is larger than the decrease in a$_{\perp}$, so the energy of the A$_1$ state slightly increases. However, at the same time strain depopulates x and y valleys and increases z-valley contribution. This reverses the change in A$_1$ due to the effect of decrease in a$_{\perp}$ being increasingly dominant on the increasingly z-valley like A$_1$ state. For $\chi < -$5, the lowest energy state is dominantly a z-valley state, with mixing from the E$_{xyz}$ state to form a new ground state A$_{1z}$. The E$_{xyz}$ state is primarily composed of x and y valleys, and therefore its energy increases with strain. It is noted that the T$_{2z}$ state does not mix with the A$_{1z}$ state as it is composed of two z-valleys with opposite signs.

\begin{figure}[t!]
\includegraphics[scale=0.3]{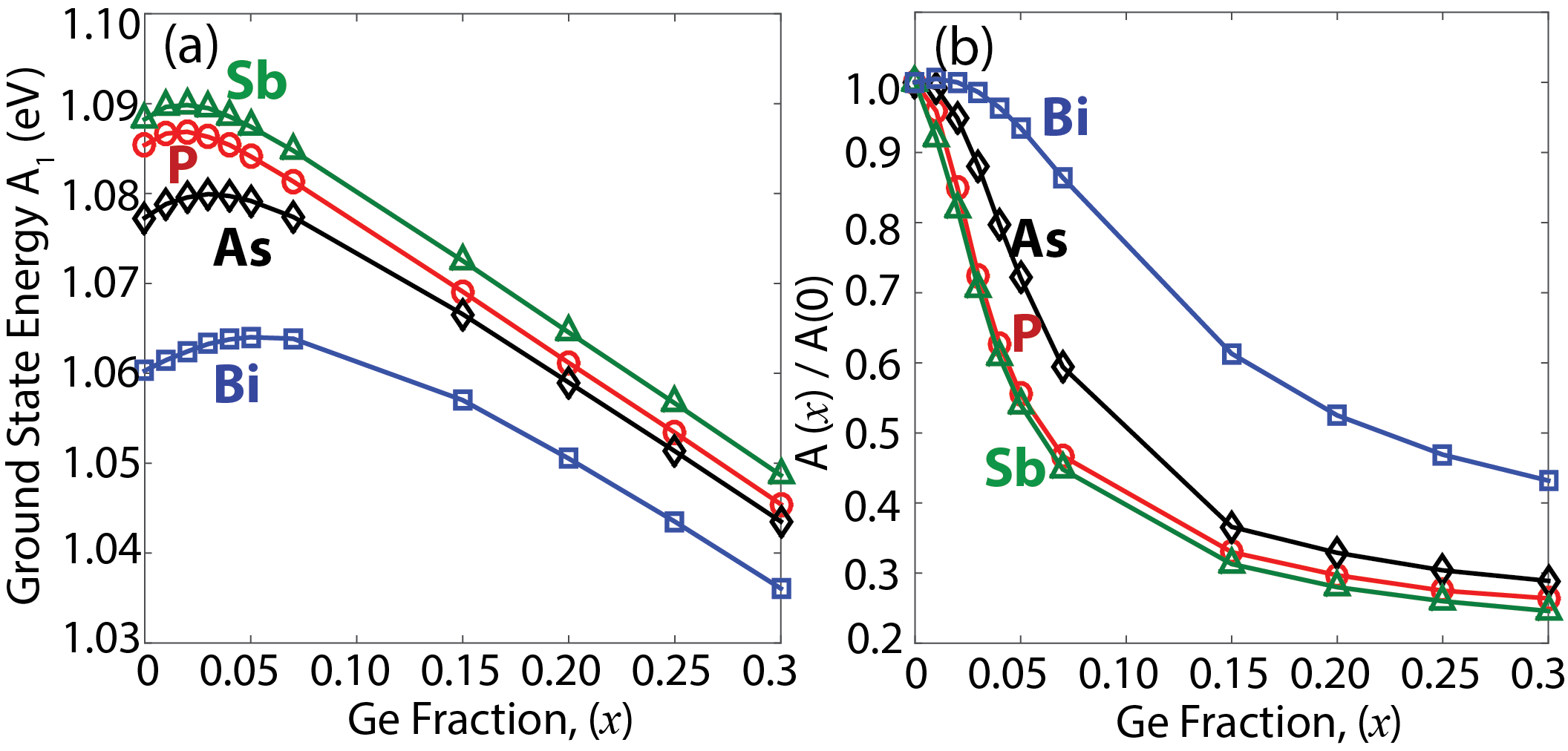}
\centering
\caption{\textbf{Theoretically predicted trends among P, Sb, As, and Bi donors :} The plots of (a) the A$_{1}$ energies and (b) the fractional change in the hyperfine couplings A($x$)/A(0) are shown as a function of the Ge content $x$.}
\label{fig:Fig2}
\end{figure}

A recent study~\cite{Pica_PRB_2014} has defined the ionization field as proportional to the energy splitting ($\delta$E) of the ground state A$_1$ and the higher excited state 2p$_0$, which is roughly 34.1 meV for P donors in bulk Si~\cite{Ramdas_RPP_1981}. As evident from Figure~\ref{fig:Fig1}(c1), the strain reduces $\delta$E, which becomes roughly 24 meV for $\chi \approx -$20, the predicted strain field to suppress the valley oscillations of the J-coupling between P donor pairs~\cite{Koiller_PRB_2002, Wellard_Hollenberg_PRB_2005}. This implies a reduction in the ionization fields for the strained P donors, which would be useful for recently proposed spin-dependent tunneling read-out schemes~\cite{McCamey_Science_2010}. The energy difference between the lowest two donor states is relevant in estimating time scales which determine the adiabatic condition in time-dependent processes driven by the gate potential variation. Figure~\ref{fig:Fig1}(c3) plots this energy difference ($\Delta$E$_{12}$ = T$_{2z} -$A$_{1z}$) as a function of $\chi$, indicating a reduction in its value due to the mixing of ground and excited states. At $\chi \approx -$20, we calculate $\Delta$E$_{12} \approx$ 2.5 meV which is smaller than the EMT value of $-$3.3 meV~\cite{Koiller_PRB_2002}.
\\ \noindent   
\textbf{3.2 Hyperfine control by strain:} \\ The hyperfine interaction A(0) is directly related to the charge density at the donor site, $|\psi$(0)$|^2$. Only the A$_1$ state has a non-zero charge density at the donor site, thus only this state contributes in the determination of A(0). The excited states T$_2$ and E do not contribute to A(0). The applied strain reduces the hyperfine coupling due to the following reasons: (1) \textit{Valley Repopulation Effect:} strain removes contributions from the x and y valleys, and increases z-valley contribution of the A$_1$ state due to mixing of the E$_{xyz}$ state. The E$_{xyz}$ state does not contribute in A(0), so A($\chi$) becomes less than A(0). (2) \textit{Crystal Deformation:} strain deforms the crystal and changes the bond-lengths from their bulk unstrained values. This modifies the radial distribution of the donor states, which are scattered over several Si lattice sites around the donor atom, leading to a reduction in hyperfine. 

The VRM model based on the EMT theory only considers the first effect and the reduction in A($\chi$) is represented by an analytical expression derived in Ref.~\cite{Wilson_PR_1961}:

\begin{equation}
	\label{eq:valley_repopulation}
	 \frac{A \left( \chi \right)}{A \left( 0 \right)}  = \frac{1}{2} \left[ 1 + \left( 1 + \frac{\chi}{6} \right) \left( 1 + \frac{\chi}{3} + \frac{\chi^2}{4} \right)^{-\frac{1}{2}} \right]  
\end{equation}

\noindent
Based on this model, the fractional change in the hyperfine A($\chi$)/A(0) as a function of $\chi$ is plotted in Figure~\ref{fig:Fig1}(d) using a dashed green line, along with the experimentally measured values (red dots) from Ref.~\citep{Huebl_PRB_2006}. Although the VRM method successfully describes A($\chi$)/A(0) for small values of strain, it fails to capture the strain effects for the larger values of the applied strain ($\chi < -$10). In fact the VRM model limits the value of A($\chi$)/A(0) to 1/3 for $\chi < -$20, based on the fact that all of the six valleys have equal contributions to the A$_1$ state, whereas only two z-valleys contribute to the A$_{1z}$ state and hence A($\chi$)/A(0)=2/6. Adding radial redistribution effects in the VRM model only reduces A($\chi$)/A(0) to 0.3 for $\chi$=$-$89~\cite{Fritzche_PR_1962}, still considerably different from the experimental data shown in Figure~\ref{fig:Fig1}(d), indicating A($\chi$)/A(0) already reduced to $\approx$ 0.22$\pm$0.09 at $\chi \approx-$29.5. Therefore in order to fully understand the strain dependence of the hyperfine, a more complete theoretical approach is required which takes into account both the valley-repopulation effect, as well as the volume deformation effect at atomistic scale. Recently reported DFT simulations~\cite{Huebl_PRB_2006} confirmed this notion by exhibiting a good match with the experimental data for both small and large values of strain (diamonds in Figure~\ref{fig:Fig1}(d)). Our TB calculations of A($\chi$)/A(0) as a function of $\chi$ are shown in the Figure~\ref{fig:Fig1}(d) using the square symbols, which demonstrate an excellent agreement with the experimental data as well as with the DFT calculations. For example, at $\chi \approx-$29.5, we calculate A($\chi$)/A(0) as $\approx$0.284, compared to the experimental value of $\approx$0.22$\pm$0.09 and the DFT value of $\approx$0.27. It is noted that the previously applied static dielectric screening of the donor potential in the TB approach~\cite{Rahman_PRL_2007, Klimeck_1, Ahmed_Enc_2009, Laucht_arxiv_2015} results in a significantly higher value of $\approx$0.364 for A($\chi$)/A(0), which emphasizes on the requirement of the non-static ($k$-dependent) screening of the donor potential for the study of the strain effects. The successful benchmarking of the TB method is in particular useful, because this approach has many advantages over the continuum EMT model and the computationally restricted DFT method. The TB theory not only accurately captures the atomistic physics, it is also scalable to simulation domains with millions of atoms, thereby enabling investigation of multi-qubit architectures. 
\\ \noindent
\textbf{3.3 Tight-binding predictions for As, Sb, and Bi donors:} \\ After benchmarking the TB theory against the experimental data set for the Si:P system, we apply it to predict the influence of strain for other three donors (As, Sb, Bi), which have drawn significant recent research interests~\cite{Bradbury_PRL_2006, Wolfowicz_NatureN_2013, Lo_arxiv_2014}. Figure~\ref{fig:Fig2}(a) plots the energies of the lowest donor state as a function of the Ge concentration $x$ in the Si$_{1-x}$Ge$_x$ substrate, which is a more relevant parameter for experimentalists compared to $\chi$ mainly used in analytical theories. Overall the changes in the A$_{1z}$ energies follow similar trends for all four donors species. Figure~\ref{fig:Fig2}(b) plots the strain-dependent hyperfine A($x$)/A(0) as a function of the Ge fraction $x$. Again the overall trends are same for all the donors: in the valley-repopulation regime, the strain-dependent hyperfine decreases sharply but for larger strain values, it becomes much less dependent on the applied strain. Interestingly, for a given value of $x$, the order of $\Delta$A($x$)=A($x$)/A(0) follows the same trend as the Stark shift parameter $\eta_2$: $\Delta$A$^{\rm Sb}$($x$) $< \Delta$A$^{\rm P}$($x$) $< \Delta$A$^{\rm As}$($x$) $< \Delta$A$^{\rm Bi}$($x$) which is same order as $\eta^{\rm Sb}_2 < \eta^{\rm P}_2 < \eta^{\rm As}_2 < \eta^{\rm Bi}_2$. One would naively expect this order to depend on the absolute values of the hyperfine interactions (A$^{\rm P}$(0) $<$ A$^{\rm Sb}$(0) $<$ A$^{\rm As}$(0) $<$ A$^{\rm Bi}$(0)), which is not true and in fact this sequence is directly related to the order of the binding energies of the donors (A$^{\rm Sb}_1 <$ A$^{\rm P}_1 <$ A$^{\rm As}_1 <$ A$^{\rm Bi}_1$).
\\ \noindent
\textbf{3.4 Benchmarking Stark shifts of hyperfine couplings against experiments:} \\ We have hitherto discussed the strain effects on the hyperfine interactions of donors, which will be useful for the psoposed all-mechanical control of qubits~\cite{Dreher_PRL_2011}. However, an alternative quantum computing architecture scheme could be based on a hybrid control of qubits, where the donors are present in strained Si and the control is applied by electric fields. Such hybrid control mechanism has certain benefits over traditional unstrained Si based systems, as the applied strain is expected to reduce valley oscillations of exchange coupling~\cite{Koiller_PRB_2002, Wellard_Hollenberg_PRB_2005}, as well as ionization fields. To study the effects of strain and electic fields simultaneously present in the qubit devices, we first calculate and benchmark the Stark shift of the hyperfine interactions for all the four donor species under study. The Stark shift is calculated by adding a potential corresponding to an electric field of magnitude varying from 0 and 0.5 MV/m in the diagonal elements of the TB Hamiltonian, and computing the hyperfine interaction A($\vec{E}$) from the field-dependent ground state A$_1$. The change in the hyperfine relative to the absolute value of the hyperfine is then fitted to a quadratic field dependence given by Eq.~\ref{eq:hyperfine_coupling}, and the Stark shift parameter ($\eta_2$) is computed by fitting to our simulation data~\cite{Rahman_PRL_2007, Usman_JPCM}:   

\begin{equation}
	\label{eq:hyperfine_coupling}
	 \frac {\textrm A \left( \vec{E} \right) - \rm{A} \left( 0 \right)}{\rm{A} \left( 0 \right)} = \eta_2 |\vec{E}|^2
\end{equation}

\begin{figure}[t!]
\centering
\includegraphics[scale=0.35]{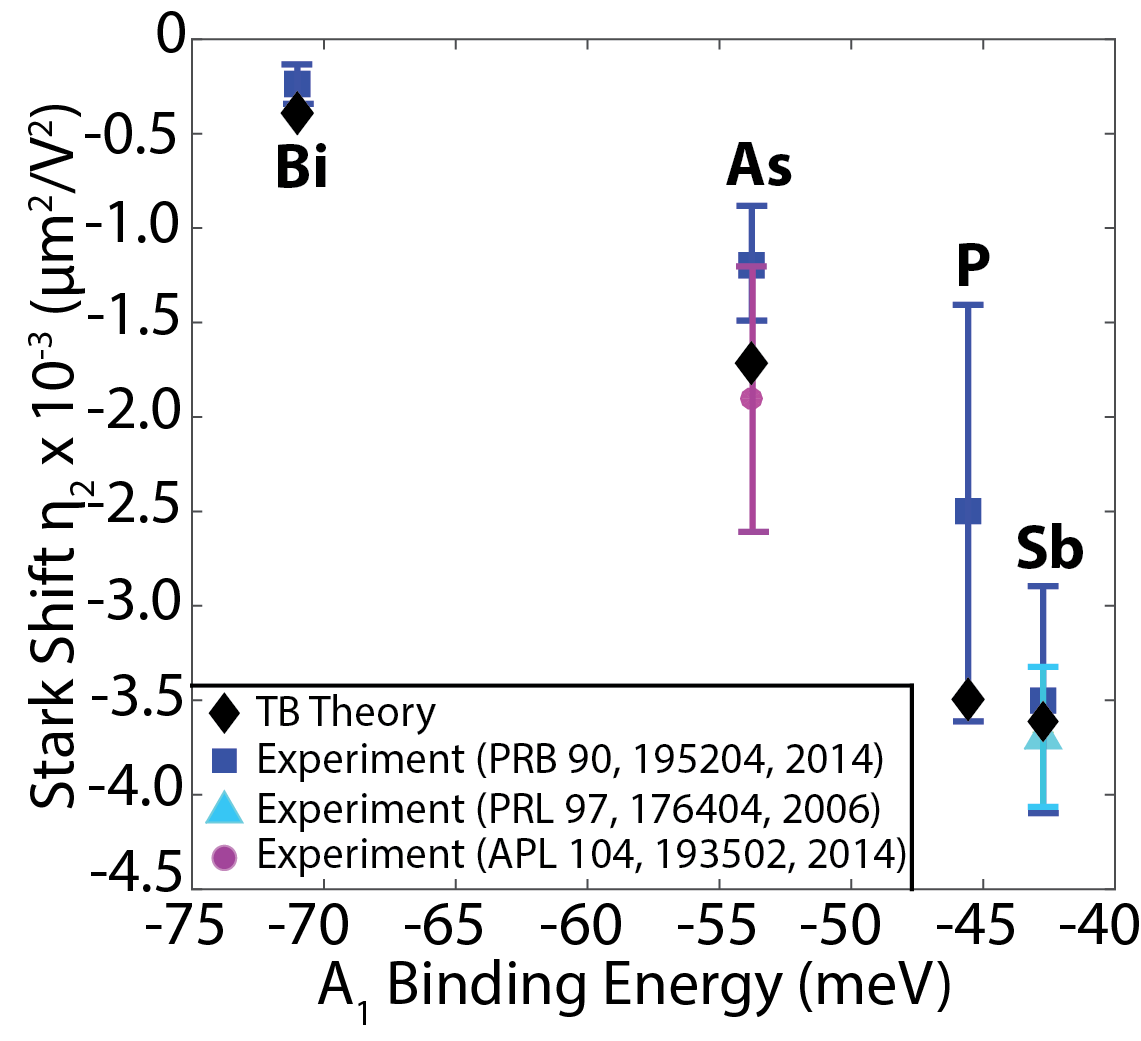}
\caption{ \textbf{Comparison of the calculated Stark shifts with experiments:} The calculated values of the Stark shift parameter $\eta_2$ are plotted as a function of the corresponding ground state binding energies (A$_1$) of shallow donors in bulk Si, which are in good agreement with the available experimental values.}
\label{fig:Fig3}
\end{figure}

\begin{figure*}[t!]
\centering
\includegraphics[scale=0.4]{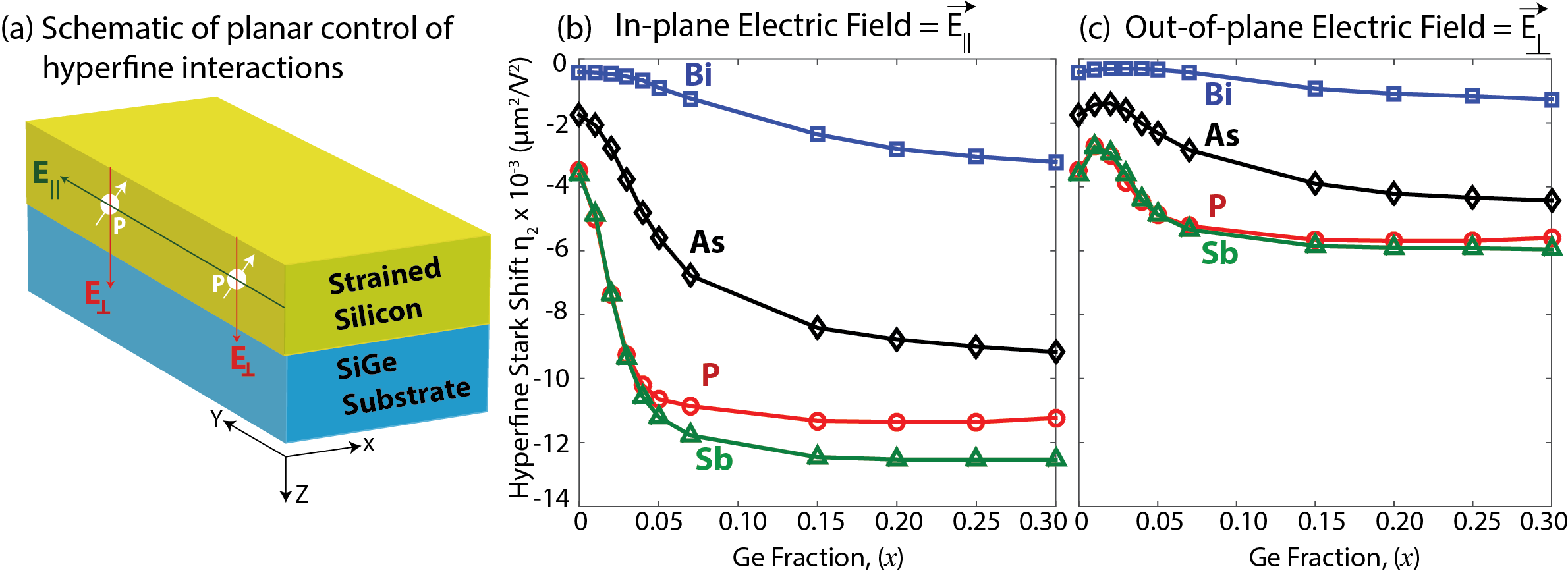}
\caption{\textbf{Planar control of hyperfine interactions in strained environments:} (a) The schematic diagram illustrates a two dimensional control of the hyperfine interaction by applying $\vec{E_{\perp}}$  and $\vec{E_{||}}$ fields along the z and y axes respectively. The plot of the strained hyperfine Stark shift $\eta_2$($x$) for (b) the in-plane field E$_{||}$ and (c) the out-of-plane field E$_{\perp}$ as a function of the applied strain ($x$). }
\label{fig:Fig4}
\end{figure*}

\noindent 
The computed values of $\eta_2$ are plotted in Figure~\ref{fig:Fig3} as a function of the A$_1$ binding energies, which are in good agreement with the available experimental measurements.
\\ \noindent
\textbf{3.5 Hybrid (Strain+Field) control of qubits:} \\ For a bulk donor in unstrained Si, the x, y, and z axes are equivalent with respect to the application of an electric field. However for the donors in strained Si, where a$_{||} \neq$ a$_{\perp}$, the effect of an electric field along the in-plane direction is expected to be different from its effect along the growth direction. Based on this notion, we suggest a new method of two-dimensional control of the donor hyperfine interaction by applying electric fields along both the growth direction (z-axis) and one of the in-plane directions (x or y-axis). Such a scheme is schematically illustrated in Figure~\ref{fig:Fig4}(a), where the P-doped strained Si is displayed on top of Si$_{1-x}$Ge$_x$ substrate. The direction of electric field can be controlled by top and side gates, which apply fields $\vec{E_{\perp}}$ and $\vec{E_{||}}$ along the growth (001) and in-plane (010) directions, respectively. In atomically precise structures, in-plane side gates are frequently used to create an electric field ($\vec{E_{||}}$) across the device~\cite{Weber_Nature_Nano_2014, Fuhrer_Nanoletters_2009} . The impact of $\vec{E_{||}}$ and $\vec{E_{\perp}}$ fields is investigated by varying their magnitudes from 0 to 0.5 MV/m, and calculating the Stark shift $\eta_2$($x$) for each value of the strain, characterized in terms of $x$.   

Figure~\ref{fig:Fig4}(b) and (c) plots the strain-dependent Stark shift values ($\eta_2$($x$)) as a function of $x$ for all of the four donor species, when the applied field is (b) $\vec{E_{||}}$ and (c) $\vec{E_{\perp}}$. In both cases, the magnitudes of $\eta_2$($x$) increase overall as a function of the strain. The increase in $|\eta_2$($x$)$|$ is larger for $\vec{E_{||}}$ compared to $\vec{E_{\perp}}$ for the same $x$. This is because of the fact that the strain compresses the spatial distribution of the donor wave function along the z-direction and therefore the effective Bohr radius along the z-axis is smaller than its values in the in-plane directions~\cite{Kettle_PRB_2006}. Consequently for the same magnitude of the electric field, the net Stark effect is stronger for the $\vec{E_{||}}$ fields than for the $\vec{E_{\perp}}$ fields. 
\\ \noindent
\textbf{3.6 Strain leads to faster gate operations:} \\ In a quantum computer placed in a static field, donors of the same species will lie at or close to resonance. The ability to Stark tune an individual, targeted, spin away from this resonance allows for the addressability of an individual qubit. However, due to a finite linewidth of an excited transition, Stark tuning by a larger frequency leads to higher fidelity and/or faster gates are achievable at the same fidelity. The timescale of an individual spin rotation is limited by the change in frequency provided by the Stark shift~\cite{Kane_Nature_1998, Pica_PRB_2014}: $\Delta f$($\vec{E}$, $x$) = $\eta_2$($x$)$\vec{E}^2$A($x$)m$\rm _I$, where m$\rm _I$ is the nuclear spin quantum number. We are interested in comparing the performance prospects of devices based on the strained Si with that of the unstrained Si, therefore we only compare the ratio of the two cases: $\Delta f$($\vec{E}$, $x$)/$\Delta f$($\vec{E}$, 0) = $\eta_2$($x$)A($x$)/$\eta_2$A(0). It is interesting to note that for the strained donors, the A($x$) value decreases (see Figure~\ref{fig:Fig2}(b)) as a function of the strain, but $|\eta_2$($x$)$|$ increases (see Figures~\ref{fig:Fig4} (b) and (c)). A multiplication of these two quantities leads to $\Delta f$($\vec{E}$, $x$)/$\Delta f$($\vec{E}$, 0) $>$ 1 for $\vec{E_{||}}$ fields, with its largest value being approximately 3 for the Bi donor. Therefore the $\vec{E_{||}}$ field allows for faster control of the qubits in the strained Si substrate in comparison to the unstrained Si. On the other hand, $\Delta f$($\vec{E}$, $x$)/$\Delta f$($\vec{E}$, 0) $<$ 1 for the $\vec{E_{\perp}}$ fields for the P, As, and Sb donors, thereby implying slower operation with the same fidelity for these donor species. Only for the Bi donor, we calculate that both the $\vec{E_{||}}$ and $\vec{E_{\perp}}$ fields exhibit $\Delta f$($\vec{E}$, $x$)/$\Delta f$($\vec{E}$, 0) $>$ 1, which is promising given that the decoherence times for Si:Bi have been reported as comparable to that of the Si:P~\cite{George_PRL_2010}. 

\section{4. Conclusions}

A multi-scale atomistic framework, established by explicitly describing the cental-cell corrections and the non-static dielectric screening of the donor potential, was used to quantitatively study the individual as well as simultaneous effects of strain and electric fields, as present in various quantum computing architecture schemes currently in development.  Our calculations were based on millions of atoms in the simulation domain, and yet described the experimentally measured strain induced hyperfine shifts with an accuracy level of the DFT method. We showed that a hybrid control scheme, where the donors are placed in a strained environment and the control of qubits is by electric fields, offers several advantages such as lowering of the ionization energies/fields and an increased magnitude of the Stark shift from the in-plane electric fields leading to the higher fidelity of single spin gates for all of the four donor species considered. The work demonstrates that the application of both strain and electric fields, and understanding their subtle interplay, has important implications for quantum control in the implementation of Si-dopant based quantum computing architectures.

\textbf{Acknowledgements:} This work is funded by the ARC Center of Excellence for Quantum Computation and Communication Technology (CE1100001027), and in part by the U.S. Army Research Office (W911NF-08-1-0527).  MYS acknowledges an ARC Laureate Fellowship. Computational resources are acknowledged from NCN/Nanohub.

\bibliographystyle{apsrev4-1}

\begin{thebibliography}{38}%
\makeatletter
\providecommand \@ifxundefined [1]{%
 \@ifx{#1\undefined}
}%
\providecommand \@ifnum [1]{%
 \ifnum #1\expandafter \@firstoftwo
 \else \expandafter \@secondoftwo
 \fi
}%
\providecommand \@ifx [1]{%
 \ifx #1\expandafter \@firstoftwo
 \else \expandafter \@secondoftwo
 \fi
}%
\providecommand \natexlab [1]{#1}%
\providecommand \enquote  [1]{``#1''}%
\providecommand \bibnamefont  [1]{#1}%
\providecommand \bibfnamefont [1]{#1}%
\providecommand \citenamefont [1]{#1}%
\providecommand \href@noop [0]{\@secondoftwo}%
\providecommand \href [0]{\begingroup \@sanitize@url \@href}%
\providecommand \@href[1]{\@@startlink{#1}\@@href}%
\providecommand \@@href[1]{\endgroup#1\@@endlink}%
\providecommand \@sanitize@url [0]{\catcode `\\12\catcode `\$12\catcode
  `\&12\catcode `\#12\catcode `\^12\catcode `\_12\catcode `\%12\relax}%
\providecommand \@@startlink[1]{}%
\providecommand \@@endlink[0]{}%
\providecommand \url  [0]{\begingroup\@sanitize@url \@url }%
\providecommand \@url [1]{\endgroup\@href {#1}{\urlprefix }}%
\providecommand \urlprefix  [0]{URL }%
\providecommand \Eprint [0]{\href }%
\providecommand \doibase [0]{http://dx.doi.org/}%
\providecommand \selectlanguage [0]{\@gobble}%
\providecommand \bibinfo  [0]{\@secondoftwo}%
\providecommand \bibfield  [0]{\@secondoftwo}%
\providecommand \translation [1]{[#1]}%
\providecommand \BibitemOpen [0]{}%
\providecommand \bibitemStop [0]{}%
\providecommand \bibitemNoStop [0]{.\EOS\space}%
\providecommand \EOS [0]{\spacefactor3000\relax}%
\providecommand \BibitemShut  [1]{\csname bibitem#1\endcsname}%
\let\auto@bib@innerbib\@empty
\bibitem [{\citenamefont {Fuechsle}\ \emph {et~al.}(2012)\citenamefont
  {Fuechsle}, \citenamefont {Miwa}, \citenamefont {Mahapatra}, \citenamefont
  {Ryu}, \citenamefont {Lee}, \citenamefont {Warschkow}, \citenamefont
  {Hollenberg}, \citenamefont {Klimeck},\ and\ \citenamefont
  {Simmons}}]{Fuechsle_NN_2012}%
  \BibitemOpen
  \bibfield  {author} {\bibinfo {author} {\bibfnamefont {M.}~\bibnamefont
  {Fuechsle}}, \bibinfo {author} {\bibfnamefont {J.~A.}\ \bibnamefont {Miwa}},
  \bibinfo {author} {\bibfnamefont {S.}~\bibnamefont {Mahapatra}}, \bibinfo
  {author} {\bibfnamefont {H.}~\bibnamefont {Ryu}}, \bibinfo {author}
  {\bibfnamefont {S.}~\bibnamefont {Lee}}, \bibinfo {author} {\bibfnamefont
  {O.}~\bibnamefont {Warschkow}}, \bibinfo {author} {\bibfnamefont {L.~C.~L.}\
  \bibnamefont {Hollenberg}}, \bibinfo {author} {\bibfnamefont
  {G.}~\bibnamefont {Klimeck}}, \ and\ \bibinfo {author} {\bibfnamefont
  {M.~Y.}\ \bibnamefont {Simmons}},\ }\href@noop {} {\bibfield  {journal}
  {\bibinfo  {journal} {Nature Nanotechnology}\ }\textbf {\bibinfo {volume}
  {7}},\ \bibinfo {pages} {242} (\bibinfo {year} {2012})}\BibitemShut {NoStop}%
\bibitem [{\citenamefont {Weber}\ \emph {et~al.}(2014)\citenamefont {Weber},
  \citenamefont {Tan}, \citenamefont {Mahapatra}, \citenamefont {Watson},
  \citenamefont {Ryu}, \citenamefont {Rahman}, \citenamefont {Hollenberg},
  \citenamefont {Klimeck},\ and\ \citenamefont
  {Simmons}}]{Weber_Nature_Nano_2014}%
  \BibitemOpen
  \bibfield  {author} {\bibinfo {author} {\bibfnamefont {B.}~\bibnamefont
  {Weber}}, \bibinfo {author} {\bibfnamefont {Y.~H.~M.}\ \bibnamefont {Tan}},
  \bibinfo {author} {\bibfnamefont {S.}~\bibnamefont {Mahapatra}}, \bibinfo
  {author} {\bibfnamefont {T.~F.}\ \bibnamefont {Watson}}, \bibinfo {author}
  {\bibfnamefont {H.}~\bibnamefont {Ryu}}, \bibinfo {author} {\bibfnamefont
  {R.}~\bibnamefont {Rahman}}, \bibinfo {author} {\bibfnamefont {L.~C.~L.}\
  \bibnamefont {Hollenberg}}, \bibinfo {author} {\bibfnamefont
  {G.}~\bibnamefont {Klimeck}}, \ and\ \bibinfo {author} {\bibfnamefont
  {M.~Y.}\ \bibnamefont {Simmons}},\ }\href@noop {} {\bibfield  {journal}
  {\bibinfo  {journal} {Nature Nanotechnology}\ }\textbf {\bibinfo {volume}
  {9}},\ \bibinfo {pages} {430} (\bibinfo {year} {2014})}\BibitemShut {NoStop}%
\bibitem [{\citenamefont {Pla}\ \emph {et~al.}(2013)\citenamefont {Pla},
  \citenamefont {Tan}, \citenamefont {Dehollain}, \citenamefont {Lim},
  \citenamefont {Morton}, \citenamefont {Zwanenburg}, \citenamefont {Jamieson},
  \citenamefont {Dzurak},\ and\ \citenamefont {Morello}}]{Pla_Nature_2013}%
  \BibitemOpen
  \bibfield  {author} {\bibinfo {author} {\bibfnamefont {J.}~\bibnamefont
  {Pla}}, \bibinfo {author} {\bibfnamefont {K.~Y.}\ \bibnamefont {Tan}},
  \bibinfo {author} {\bibfnamefont {J.~P.}\ \bibnamefont {Dehollain}}, \bibinfo
  {author} {\bibfnamefont {W.~H.}\ \bibnamefont {Lim}}, \bibinfo {author}
  {\bibfnamefont {J.~J.~L.}\ \bibnamefont {Morton}}, \bibinfo {author}
  {\bibfnamefont {F.~A.}\ \bibnamefont {Zwanenburg}}, \bibinfo {author}
  {\bibfnamefont {D.~N.}\ \bibnamefont {Jamieson}}, \bibinfo {author}
  {\bibfnamefont {A.~S.}\ \bibnamefont {Dzurak}}, \ and\ \bibinfo {author}
  {\bibfnamefont {A.}~\bibnamefont {Morello}},\ }\href@noop {} {\bibfield
  {journal} {\bibinfo  {journal} {Nature}\ }\textbf {\bibinfo {volume} {496}},\
  \bibinfo {pages} {334} (\bibinfo {year} {2013})}\BibitemShut {NoStop}%
\bibitem [{\citenamefont {Saeedi}\ \emph {et~al.}(2013)\citenamefont {Saeedi},
  \citenamefont {Simmons}, \citenamefont {Salvail}, \citenamefont {Dluhy},
  \citenamefont {Riemann}, \citenamefont {Abrosimov}, \citenamefont {Becker},
  \citenamefont {Pohl}, \citenamefont {Morton},\ and\ \citenamefont
  {Thewalt}}]{Saeedi_Science_2013}%
  \BibitemOpen
  \bibfield  {author} {\bibinfo {author} {\bibfnamefont {K.}~\bibnamefont
  {Saeedi}}, \bibinfo {author} {\bibfnamefont {S.}~\bibnamefont {Simmons}},
  \bibinfo {author} {\bibfnamefont {J.~Z.}\ \bibnamefont {Salvail}}, \bibinfo
  {author} {\bibfnamefont {P.}~\bibnamefont {Dluhy}}, \bibinfo {author}
  {\bibfnamefont {H.}~\bibnamefont {Riemann}}, \bibinfo {author} {\bibfnamefont
  {N.~V.}\ \bibnamefont {Abrosimov}}, \bibinfo {author} {\bibfnamefont
  {P.}~\bibnamefont {Becker}}, \bibinfo {author} {\bibfnamefont {H.-J.}\
  \bibnamefont {Pohl}}, \bibinfo {author} {\bibfnamefont {J.~J.~L.}\
  \bibnamefont {Morton}}, \ and\ \bibinfo {author} {\bibfnamefont {M.~L.~W.}\
  \bibnamefont {Thewalt}},\ }\href@noop {} {\bibfield  {journal} {\bibinfo
  {journal} {Science}\ }\textbf {\bibinfo {volume} {342}},\ \bibinfo {pages}
  {830} (\bibinfo {year} {2013})}\BibitemShut {NoStop}%
\bibitem [{\citenamefont {Muhonen}\ \emph {et~al.}(2014)\citenamefont
  {Muhonen}, \citenamefont {Dehollain}, \citenamefont {Laucht}, \citenamefont
  {Hudson}, \citenamefont {Kalra}, \citenamefont {Sekiguchi}, \citenamefont
  {Itoh}, \citenamefont {Jamieson}, \citenamefont {McCallum}, \citenamefont
  {Dzurak},\ and\ \citenamefont {Morello}}]{Muhonen_NatureNano_2014}%
  \BibitemOpen
  \bibfield  {author} {\bibinfo {author} {\bibfnamefont {J.~T.}\ \bibnamefont
  {Muhonen}}, \bibinfo {author} {\bibfnamefont {J.~P.}\ \bibnamefont
  {Dehollain}}, \bibinfo {author} {\bibfnamefont {A.}~\bibnamefont {Laucht}},
  \bibinfo {author} {\bibfnamefont {F.~E.}\ \bibnamefont {Hudson}}, \bibinfo
  {author} {\bibfnamefont {R.}~\bibnamefont {Kalra}}, \bibinfo {author}
  {\bibfnamefont {T.}~\bibnamefont {Sekiguchi}}, \bibinfo {author}
  {\bibfnamefont {K.~M.}\ \bibnamefont {Itoh}}, \bibinfo {author}
  {\bibfnamefont {D.~N.}\ \bibnamefont {Jamieson}}, \bibinfo {author}
  {\bibfnamefont {J.~C.}\ \bibnamefont {McCallum}}, \bibinfo {author}
  {\bibfnamefont {A.~S.}\ \bibnamefont {Dzurak}}, \ and\ \bibinfo {author}
  {\bibfnamefont {A.}~\bibnamefont {Morello}},\ }\href@noop {} {\bibfield
  {journal} {\bibinfo  {journal} {Nature Nanotechnology}\ }\textbf {\bibinfo
  {volume} {9}},\ \bibinfo {pages} {986} (\bibinfo {year} {2014})}\BibitemShut
  {NoStop}%
\bibitem [{\citenamefont {Kane}(1998)}]{Kane_Nature_1998}%
  \BibitemOpen
  \bibfield  {author} {\bibinfo {author} {\bibfnamefont {B.~E.}\ \bibnamefont
  {Kane}},\ }\href@noop {} {\bibfield  {journal} {\bibinfo  {journal} {Nature}\
  }\textbf {\bibinfo {volume} {393}},\ \bibinfo {pages} {133} (\bibinfo {year}
  {1998})}\BibitemShut {NoStop}%
\bibitem [{\citenamefont {Hollenberg}\ \emph {et~al.}(2006)\citenamefont
  {Hollenberg}, \citenamefont {Greentree}, \citenamefont {Fowler},\ and\
  \citenamefont {Wellard}}]{Hollenberg_PRB_2006}%
  \BibitemOpen
  \bibfield  {author} {\bibinfo {author} {\bibfnamefont {L.}~\bibnamefont
  {Hollenberg}}, \bibinfo {author} {\bibfnamefont {A.~D.}\ \bibnamefont
  {Greentree}}, \bibinfo {author} {\bibfnamefont {A.~G.}\ \bibnamefont
  {Fowler}}, \ and\ \bibinfo {author} {\bibfnamefont {C.~J.}\ \bibnamefont
  {Wellard}},\ }\href@noop {} {\bibfield  {journal} {\bibinfo  {journal} {Phys.
  Rev. B}\ }\textbf {\bibinfo {volume} {74}},\ \bibinfo {pages} {045311}
  (\bibinfo {year} {2006})}\BibitemShut {NoStop}%
\bibitem [{\citenamefont {Lansbergen}\ \emph {et~al.}(2008)\citenamefont
  {Lansbergen}, \citenamefont {Rahman}, \citenamefont {Wellard}, \citenamefont
  {Woo}, \citenamefont {Caro}, \citenamefont {Collaert}, \citenamefont
  {Biesemans}, \citenamefont {Klimeck}, \citenamefont {Hollenberg},\ and\
  \citenamefont {Rogge}}]{Lansbergen_Nat_Phys_2008}%
  \BibitemOpen
  \bibfield  {author} {\bibinfo {author} {\bibfnamefont {G.~P.}\ \bibnamefont
  {Lansbergen}}, \bibinfo {author} {\bibfnamefont {R.}~\bibnamefont {Rahman}},
  \bibinfo {author} {\bibfnamefont {C.~J.}\ \bibnamefont {Wellard}}, \bibinfo
  {author} {\bibfnamefont {I.}~\bibnamefont {Woo}}, \bibinfo {author}
  {\bibfnamefont {J.}~\bibnamefont {Caro}}, \bibinfo {author} {\bibfnamefont
  {N.}~\bibnamefont {Collaert}}, \bibinfo {author} {\bibfnamefont
  {S.}~\bibnamefont {Biesemans}}, \bibinfo {author} {\bibfnamefont
  {G.}~\bibnamefont {Klimeck}}, \bibinfo {author} {\bibfnamefont {L.~C.~L.}\
  \bibnamefont {Hollenberg}}, \ and\ \bibinfo {author} {\bibfnamefont
  {S.}~\bibnamefont {Rogge}},\ }\href@noop {} {\bibfield  {journal} {\bibinfo
  {journal} {Nature Physics}\ }\textbf {\bibinfo {volume} {4}},\ \bibinfo
  {pages} {656} (\bibinfo {year} {2008})}\BibitemShut {NoStop}%
\bibitem [{\citenamefont {Huebl}\ \emph {et~al.}(2006)\citenamefont {Huebl},
  \citenamefont {Stegner}, \citenamefont {Stutzmann}, \citenamefont {Brandt},
  \citenamefont {Vogg}, \citenamefont {Bensch}, \citenamefont {Rauls},\ and\
  \citenamefont {Gerstmann}}]{Huebl_PRB_2006}%
  \BibitemOpen
  \bibfield  {author} {\bibinfo {author} {\bibfnamefont {H.}~\bibnamefont
  {Huebl}}, \bibinfo {author} {\bibfnamefont {A.~R.}\ \bibnamefont {Stegner}},
  \bibinfo {author} {\bibfnamefont {M.}~\bibnamefont {Stutzmann}}, \bibinfo
  {author} {\bibfnamefont {M.~S.}\ \bibnamefont {Brandt}}, \bibinfo {author}
  {\bibfnamefont {G.}~\bibnamefont {Vogg}}, \bibinfo {author} {\bibfnamefont
  {F.}~\bibnamefont {Bensch}}, \bibinfo {author} {\bibfnamefont
  {E.}~\bibnamefont {Rauls}}, \ and\ \bibinfo {author} {\bibfnamefont
  {U.}~\bibnamefont {Gerstmann}},\ }\href@noop {} {\bibfield  {journal}
  {\bibinfo  {journal} {Phys. Rev. Lett.}\ }\textbf {\bibinfo {volume} {97}},\
  \bibinfo {pages} {166402} (\bibinfo {year} {2006})}\BibitemShut {NoStop}%
\bibitem [{\citenamefont {Dreher}\ \emph {et~al.}(2011)\citenamefont {Dreher},
  \citenamefont {Hilker}, \citenamefont {Brandlmaier}, \citenamefont
  {Goennenwein}, \citenamefont {Huebl}, \citenamefont {Stutzmann},\ and\
  \citenamefont {Brandt}}]{Dreher_PRL_2011}%
  \BibitemOpen
  \bibfield  {author} {\bibinfo {author} {\bibfnamefont {L.}~\bibnamefont
  {Dreher}}, \bibinfo {author} {\bibfnamefont {T.~A.}\ \bibnamefont {Hilker}},
  \bibinfo {author} {\bibfnamefont {A.}~\bibnamefont {Brandlmaier}}, \bibinfo
  {author} {\bibfnamefont {S.~T.~B.}\ \bibnamefont {Goennenwein}}, \bibinfo
  {author} {\bibfnamefont {H.}~\bibnamefont {Huebl}}, \bibinfo {author}
  {\bibfnamefont {M.}~\bibnamefont {Stutzmann}}, \ and\ \bibinfo {author}
  {\bibfnamefont {M.~S.}\ \bibnamefont {Brandt}},\ }\href@noop {} {\bibfield
  {journal} {\bibinfo  {journal} {Phys. Rev. Lett.}\ }\textbf {\bibinfo
  {volume} {106}},\ \bibinfo {pages} {037601} (\bibinfo {year}
  {2011})}\BibitemShut {NoStop}%
\bibitem [{\citenamefont {Martins}\ \emph {et~al.}(2004)\citenamefont
  {Martins}, \citenamefont {Capaz},\ and\ \citenamefont
  {Koiller}}]{Martins_PRB_2004}%
  \BibitemOpen
  \bibfield  {author} {\bibinfo {author} {\bibfnamefont {A.~S.}\ \bibnamefont
  {Martins}}, \bibinfo {author} {\bibfnamefont {R.~B.}\ \bibnamefont {Capaz}},
  \ and\ \bibinfo {author} {\bibfnamefont {B.}~\bibnamefont {Koiller}},\
  }\href@noop {} {\bibfield  {journal} {\bibinfo  {journal} {Phys. Rev. B}\
  }\textbf {\bibinfo {volume} {69}},\ \bibinfo {pages} {085320} (\bibinfo
  {year} {2004})}\BibitemShut {NoStop}%
\bibitem [{\citenamefont {Rahman}\ \emph {et~al.}(2007)\citenamefont {Rahman},
  \citenamefont {Wellard}, \citenamefont {Bradbury}, \citenamefont {Prada},
  \citenamefont {Cole}, \citenamefont {Klimeck},\ and\ \citenamefont
  {Hollenberg}}]{Rahman_PRL_2007}%
  \BibitemOpen
  \bibfield  {author} {\bibinfo {author} {\bibfnamefont {R.}~\bibnamefont
  {Rahman}}, \bibinfo {author} {\bibfnamefont {C.~J.}\ \bibnamefont {Wellard}},
  \bibinfo {author} {\bibfnamefont {F.~R.}\ \bibnamefont {Bradbury}}, \bibinfo
  {author} {\bibfnamefont {M.}~\bibnamefont {Prada}}, \bibinfo {author}
  {\bibfnamefont {J.~H.}\ \bibnamefont {Cole}}, \bibinfo {author}
  {\bibfnamefont {G.}~\bibnamefont {Klimeck}}, \ and\ \bibinfo {author}
  {\bibfnamefont {L.~C.~L.}\ \bibnamefont {Hollenberg}},\ }\href@noop {}
  {\bibfield  {journal} {\bibinfo  {journal} {Phys. Rev. Lett.}\ }\textbf
  {\bibinfo {volume} {99}},\ \bibinfo {pages} {036403} (\bibinfo {year}
  {2007})}\BibitemShut {NoStop}%
\bibitem [{\citenamefont {Usman}\ \emph {et~al.}(2015)\citenamefont {Usman},
  \citenamefont {Rahman}, \citenamefont {Salfi}, \citenamefont {Bocquel},
  \citenamefont {Voisin}, \citenamefont {Rogge}, \citenamefont {Klimeck},\ and\
  \citenamefont {Hollenberg}}]{Usman_JPCM}%
  \BibitemOpen
  \bibfield  {author} {\bibinfo {author} {\bibfnamefont {M.}~\bibnamefont
  {Usman}}, \bibinfo {author} {\bibfnamefont {R.}~\bibnamefont {Rahman}},
  \bibinfo {author} {\bibfnamefont {J.}~\bibnamefont {Salfi}}, \bibinfo
  {author} {\bibfnamefont {J.}~\bibnamefont {Bocquel}}, \bibinfo {author}
  {\bibfnamefont {B.}~\bibnamefont {Voisin}}, \bibinfo {author} {\bibfnamefont
  {S.}~\bibnamefont {Rogge}}, \bibinfo {author} {\bibfnamefont
  {G.}~\bibnamefont {Klimeck}}, \ and\ \bibinfo {author} {\bibfnamefont
  {L.~C.~L.}\ \bibnamefont {Hollenberg}},\ }\href@noop {} {\bibfield  {journal}
  {\bibinfo  {journal} {J. Phys.: Cond. Matt.}\ }\textbf {\bibinfo {volume}
  {27}},\ \bibinfo {pages} {154207} (\bibinfo {year} {2015})}\BibitemShut
  {NoStop}%
\bibitem [{\citenamefont {Pica}\ \emph {et~al.}(2014)\citenamefont {Pica},
  \citenamefont {Wolfowicz}, \citenamefont {Urdampilleta}, \citenamefont
  {Thewalt}, \citenamefont {Riemann}, \citenamefont {Abrosimov}, \citenamefont
  {Becker}, \citenamefont {Pohl}, \citenamefont {Morton}, \citenamefont
  {Bhatt}, \citenamefont {Lyon},\ and\ \citenamefont {Lovett}}]{Pica_PRB_2014}%
  \BibitemOpen
  \bibfield  {author} {\bibinfo {author} {\bibfnamefont {G.}~\bibnamefont
  {Pica}}, \bibinfo {author} {\bibfnamefont {G.}~\bibnamefont {Wolfowicz}},
  \bibinfo {author} {\bibfnamefont {M.}~\bibnamefont {Urdampilleta}}, \bibinfo
  {author} {\bibfnamefont {M.~L.~W.}\ \bibnamefont {Thewalt}}, \bibinfo
  {author} {\bibfnamefont {H.}~\bibnamefont {Riemann}}, \bibinfo {author}
  {\bibfnamefont {N.~V.}\ \bibnamefont {Abrosimov}}, \bibinfo {author}
  {\bibfnamefont {P.}~\bibnamefont {Becker}}, \bibinfo {author} {\bibfnamefont
  {H.-J.}\ \bibnamefont {Pohl}}, \bibinfo {author} {\bibfnamefont {J.~J.~L.}\
  \bibnamefont {Morton}}, \bibinfo {author} {\bibfnamefont {R.~N.}\
  \bibnamefont {Bhatt}}, \bibinfo {author} {\bibfnamefont {S.~A.}\ \bibnamefont
  {Lyon}}, \ and\ \bibinfo {author} {\bibfnamefont {B.~W.}\ \bibnamefont
  {Lovett}},\ }\href@noop {} {\bibfield  {journal} {\bibinfo  {journal} {Phys.
  Rev. B}\ }\textbf {\bibinfo {volume} {90}},\ \bibinfo {pages} {195204}
  (\bibinfo {year} {2014})}\BibitemShut {NoStop}%
\bibitem [{\citenamefont {Zwanenburg}\ \emph {et~al.}(2013)\citenamefont
  {Zwanenburg}, \citenamefont {Dzurak}, \citenamefont {Morello}, \citenamefont
  {Simmons}, \citenamefont {Hollenberg}, \citenamefont {Klimeck}, \citenamefont
  {Rogge}, \citenamefont {Coppersmith},\ and\ \citenamefont
  {Eriksson}}]{Zwanenburg_RMP_2013}%
  \BibitemOpen
  \bibfield  {author} {\bibinfo {author} {\bibfnamefont {F.}~\bibnamefont
  {Zwanenburg}}, \bibinfo {author} {\bibfnamefont {A.~S.}\ \bibnamefont
  {Dzurak}}, \bibinfo {author} {\bibfnamefont {A.}~\bibnamefont {Morello}},
  \bibinfo {author} {\bibfnamefont {M.~Y.}\ \bibnamefont {Simmons}}, \bibinfo
  {author} {\bibfnamefont {L.~C.~L.}\ \bibnamefont {Hollenberg}}, \bibinfo
  {author} {\bibfnamefont {G.}~\bibnamefont {Klimeck}}, \bibinfo {author}
  {\bibfnamefont {S.}~\bibnamefont {Rogge}}, \bibinfo {author} {\bibfnamefont
  {S.~N.}\ \bibnamefont {Coppersmith}}, \ and\ \bibinfo {author} {\bibfnamefont
  {M.~A.}\ \bibnamefont {Eriksson}},\ }\href@noop {} {\bibfield  {journal}
  {\bibinfo  {journal} {Rev. Mod. Phys.}\ }\textbf {\bibinfo {volume} {85}},\
  \bibinfo {pages} {961} (\bibinfo {year} {2013})}\BibitemShut {NoStop}%
\bibitem [{\citenamefont {Bradbury}\ \emph {et~al.}(2006)\citenamefont
  {Bradbury}, \citenamefont {Tyryshkin}, \citenamefont {Sabouret},
  \citenamefont {Bokor}, \citenamefont {Schenkel},\ and\ \citenamefont
  {Lyon}}]{Bradbury_PRL_2006}%
  \BibitemOpen
  \bibfield  {author} {\bibinfo {author} {\bibfnamefont {F.~R.}\ \bibnamefont
  {Bradbury}}, \bibinfo {author} {\bibfnamefont {A.~M.}\ \bibnamefont
  {Tyryshkin}}, \bibinfo {author} {\bibfnamefont {G.}~\bibnamefont {Sabouret}},
  \bibinfo {author} {\bibfnamefont {J.}~\bibnamefont {Bokor}}, \bibinfo
  {author} {\bibfnamefont {T.}~\bibnamefont {Schenkel}}, \ and\ \bibinfo
  {author} {\bibfnamefont {S.~A.}\ \bibnamefont {Lyon}},\ }\href@noop {}
  {\bibfield  {journal} {\bibinfo  {journal} {Phys. Rev. Lett.}\ }\textbf
  {\bibinfo {volume} {97}},\ \bibinfo {pages} {176404} (\bibinfo {year}
  {2006})}\BibitemShut {NoStop}%
\bibitem [{\citenamefont {Lo}\ \emph {et~al.}(2014)\citenamefont {Lo},
  \citenamefont {Simmons}, \citenamefont {Nardo}, \citenamefont {Weis},
  \citenamefont {Tyryshkin}, \citenamefont {Meijer}, \citenamefont {Rogalla},
  \citenamefont {Lyon}, \citenamefont {Bokor}, \citenamefont {Schenkel},\ and\
  \citenamefont {Morton1}}]{Lo_arxiv_2014}%
  \BibitemOpen
  \bibfield  {author} {\bibinfo {author} {\bibfnamefont {C.~C.}\ \bibnamefont
  {Lo}}, \bibinfo {author} {\bibfnamefont {S.}~\bibnamefont {Simmons}},
  \bibinfo {author} {\bibfnamefont {R.~L.}\ \bibnamefont {Nardo}}, \bibinfo
  {author} {\bibfnamefont {C.~D.}\ \bibnamefont {Weis}}, \bibinfo {author}
  {\bibfnamefont {A.~M.}\ \bibnamefont {Tyryshkin}}, \bibinfo {author}
  {\bibfnamefont {J.}~\bibnamefont {Meijer}}, \bibinfo {author} {\bibfnamefont
  {D.}~\bibnamefont {Rogalla}}, \bibinfo {author} {\bibfnamefont {S.~A.}\
  \bibnamefont {Lyon}}, \bibinfo {author} {\bibfnamefont {J.}~\bibnamefont
  {Bokor}}, \bibinfo {author} {\bibfnamefont {T.}~\bibnamefont {Schenkel}}, \
  and\ \bibinfo {author} {\bibfnamefont {J.~J.~L.}\ \bibnamefont {Morton1}},\
  }\href@noop {} {\bibfield  {journal} {\bibinfo  {journal} {Appl. Phys.
  Lett.}\ }\textbf {\bibinfo {volume} {104}},\ \bibinfo {pages} {193502}
  (\bibinfo {year} {2014})}\BibitemShut {NoStop}%
\bibitem [{\citenamefont {Koiller}\ \emph {et~al.}(2002)\citenamefont
  {Koiller}, \citenamefont {Hu},\ and\ \citenamefont
  {Sarma}}]{Koiller_PRB_2002}%
  \BibitemOpen
  \bibfield  {author} {\bibinfo {author} {\bibfnamefont {B.}~\bibnamefont
  {Koiller}}, \bibinfo {author} {\bibfnamefont {X.}~\bibnamefont {Hu}}, \ and\
  \bibinfo {author} {\bibfnamefont {S.~D.}\ \bibnamefont {Sarma}},\ }\href@noop
  {} {\bibfield  {journal} {\bibinfo  {journal} {Phys. Rev. B}\ }\textbf
  {\bibinfo {volume} {66}},\ \bibinfo {pages} {115201} (\bibinfo {year}
  {2002})}\BibitemShut {NoStop}%
\bibitem [{\citenamefont {Wellard}\ and\ \citenamefont
  {Hollenberg}(2005)}]{Wellard_Hollenberg_PRB_2005}%
  \BibitemOpen
  \bibfield  {author} {\bibinfo {author} {\bibfnamefont {C.~J.}\ \bibnamefont
  {Wellard}}\ and\ \bibinfo {author} {\bibfnamefont {L.~C.~L.}\ \bibnamefont
  {Hollenberg}},\ }\href@noop {} {\bibfield  {journal} {\bibinfo  {journal}
  {Phys. Rev. B}\ }\textbf {\bibinfo {volume} {72}},\ \bibinfo {pages} {085202}
  (\bibinfo {year} {2005})}\BibitemShut {NoStop}%
\bibitem [{\citenamefont {Lee}\ \emph {et~al.}(2014)\citenamefont {Lee},
  \citenamefont {McKibbin}, \citenamefont {Thompson}, \citenamefont {Xue},
  \citenamefont {Scappucci}, \citenamefont {Bishop}, \citenamefont {Celler},
  \citenamefont {Carroll},\ and\ \citenamefont {Simmons}}]{Lee_Nanotech_2014}%
  \BibitemOpen
  \bibfield  {author} {\bibinfo {author} {\bibfnamefont {W.}~\bibnamefont
  {Lee}}, \bibinfo {author} {\bibfnamefont {S.~R.}\ \bibnamefont {McKibbin}},
  \bibinfo {author} {\bibfnamefont {D.~L.}\ \bibnamefont {Thompson}}, \bibinfo
  {author} {\bibfnamefont {K.}~\bibnamefont {Xue}}, \bibinfo {author}
  {\bibfnamefont {G.}~\bibnamefont {Scappucci}}, \bibinfo {author}
  {\bibfnamefont {N.}~\bibnamefont {Bishop}}, \bibinfo {author} {\bibfnamefont
  {G.~K.}\ \bibnamefont {Celler}}, \bibinfo {author} {\bibfnamefont {M.~S.}\
  \bibnamefont {Carroll}}, \ and\ \bibinfo {author} {\bibfnamefont {M.~Y.}\
  \bibnamefont {Simmons}},\ }\href@noop {} {\bibfield  {journal} {\bibinfo
  {journal} {Nanotechnology}\ }\textbf {\bibinfo {volume} {25}},\ \bibinfo
  {pages} {145302} (\bibinfo {year} {2014})}\BibitemShut {NoStop}%
\bibitem [{\citenamefont {Wilson}\ and\ \citenamefont
  {Feher}(1961)}]{Wilson_PR_1961}%
  \BibitemOpen
  \bibfield  {author} {\bibinfo {author} {\bibfnamefont {D.~K.}\ \bibnamefont
  {Wilson}}\ and\ \bibinfo {author} {\bibfnamefont {G.}~\bibnamefont {Feher}},\
  }\href@noop {} {\bibfield  {journal} {\bibinfo  {journal} {Phys. Rev.}\
  }\textbf {\bibinfo {volume} {124}},\ \bibinfo {pages} {1068} (\bibinfo {year}
  {1961})}\BibitemShut {NoStop}%
\bibitem [{\citenamefont {Fritzsche}(1962)}]{Fritzche_PR_1962}%
  \BibitemOpen
  \bibfield  {author} {\bibinfo {author} {\bibfnamefont {H.}~\bibnamefont
  {Fritzsche}},\ }\href@noop {} {\bibfield  {journal} {\bibinfo  {journal}
  {Phys. Rev.}\ }\textbf {\bibinfo {volume} {125}},\ \bibinfo {pages} {1560}
  (\bibinfo {year} {1962})}\BibitemShut {NoStop}%
\bibitem [{\citenamefont {Hale}\ and\ \citenamefont {Castner}(1970)}]{Hale_PR}%
  \BibitemOpen
  \bibfield  {author} {\bibinfo {author} {\bibfnamefont {E.~B.}\ \bibnamefont
  {Hale}}\ and\ \bibinfo {author} {\bibfnamefont {T.~G.}\ \bibnamefont
  {Castner}},\ }\href@noop {} {\bibfield  {journal} {\bibinfo  {journal} {Phys.
  Rev. B}\ }\textbf {\bibinfo {volume} {1}},\ \bibinfo {pages} {4763} (\bibinfo
  {year} {1970})}\BibitemShut {NoStop}%
\bibitem [{\citenamefont {Overhof}\ and\ \citenamefont
  {Gerstmann}(2004)}]{Overhof_PRL_2004}%
  \BibitemOpen
  \bibfield  {author} {\bibinfo {author} {\bibfnamefont {H.}~\bibnamefont
  {Overhof}}\ and\ \bibinfo {author} {\bibfnamefont {U.}~\bibnamefont
  {Gerstmann}},\ }\href@noop {} {\bibfield  {journal} {\bibinfo  {journal}
  {Phys. Rev. Lett.}\ }\textbf {\bibinfo {volume} {92}},\ \bibinfo {pages}
  {087602} (\bibinfo {year} {2004})}\BibitemShut {NoStop}%
\bibitem [{\citenamefont {Klimeck}\ \emph
  {et~al.}(2007{\natexlab{a}})\citenamefont {Klimeck}, \citenamefont {Ahmed},
  \citenamefont {Bae}, \citenamefont {Kharche}, \citenamefont {Clark},
  \citenamefont {Haley}, \citenamefont {Lee}, \citenamefont {Naumov},
  \citenamefont {Ryu}, \citenamefont {Saied}, \citenamefont {.Prada},
  \citenamefont {Korkusinski}, \citenamefont {Boykin},\ and\ \citenamefont
  {Rahman}}]{Klimeck_1}%
  \BibitemOpen
  \bibfield  {author} {\bibinfo {author} {\bibfnamefont {G.}~\bibnamefont
  {Klimeck}}, \bibinfo {author} {\bibfnamefont {S.}~\bibnamefont {Ahmed}},
  \bibinfo {author} {\bibfnamefont {H.}~\bibnamefont {Bae}}, \bibinfo {author}
  {\bibfnamefont {N.}~\bibnamefont {Kharche}}, \bibinfo {author} {\bibfnamefont
  {S.}~\bibnamefont {Clark}}, \bibinfo {author} {\bibfnamefont
  {B.}~\bibnamefont {Haley}}, \bibinfo {author} {\bibfnamefont
  {S.}~\bibnamefont {Lee}}, \bibinfo {author} {\bibfnamefont {M.}~\bibnamefont
  {Naumov}}, \bibinfo {author} {\bibfnamefont {H.}~\bibnamefont {Ryu}},
  \bibinfo {author} {\bibfnamefont {F.}~\bibnamefont {Saied}}, \bibinfo
  {author} {\bibfnamefont {M.}~\bibnamefont {.Prada}}, \bibinfo {author}
  {\bibfnamefont {M.}~\bibnamefont {Korkusinski}}, \bibinfo {author}
  {\bibfnamefont {T.~B.}\ \bibnamefont {Boykin}}, \ and\ \bibinfo {author}
  {\bibfnamefont {R.}~\bibnamefont {Rahman}},\ }\href@noop {} {\bibfield
  {journal} {\bibinfo  {journal} {IEEE Trans. Elect. Dev.}\ }\textbf {\bibinfo
  {volume} {54}},\ \bibinfo {pages} {2079} (\bibinfo {year}
  {2007}{\natexlab{a}})}\BibitemShut {NoStop}%
\bibitem [{\citenamefont {Klimeck}\ \emph
  {et~al.}(2007{\natexlab{b}})\citenamefont {Klimeck}, \citenamefont {Ahmed},
  \citenamefont {Kharche}, \citenamefont {Korkusinski}, \citenamefont {Usman},
  \citenamefont {Parada},\ and\ \citenamefont {Boykin}}]{Klimeck_2}%
  \BibitemOpen
  \bibfield  {author} {\bibinfo {author} {\bibfnamefont {G.}~\bibnamefont
  {Klimeck}}, \bibinfo {author} {\bibfnamefont {S.}~\bibnamefont {Ahmed}},
  \bibinfo {author} {\bibfnamefont {N.}~\bibnamefont {Kharche}}, \bibinfo
  {author} {\bibfnamefont {M.}~\bibnamefont {Korkusinski}}, \bibinfo {author}
  {\bibfnamefont {M.}~\bibnamefont {Usman}}, \bibinfo {author} {\bibfnamefont
  {M.}~\bibnamefont {Parada}}, \ and\ \bibinfo {author} {\bibfnamefont
  {T.}~\bibnamefont {Boykin}},\ }\href@noop {} {\bibfield  {journal} {\bibinfo
  {journal} {IEEE Trans. Elect. Dev.}\ }\textbf {\bibinfo {volume} {54}},\
  \bibinfo {pages} {2090} (\bibinfo {year} {2007}{\natexlab{b}})}\BibitemShut
  {NoStop}%
\bibitem [{\citenamefont {Laucht}\ \emph {et~al.}(2015)\citenamefont {Laucht},
  \citenamefont {Muhonen}, \citenamefont {Mohiyaddin}, \citenamefont {Kalra},
  \citenamefont {Dehollain}, \citenamefont {Freer}, \citenamefont {Hudson},
  \citenamefont {Veldhorst}, \citenamefont {Rahman}, \citenamefont {Klimeck},
  \citenamefont {Itoh}, \citenamefont {Jamieson}, \citenamefont {McCallum},
  \citenamefont {Dzurak},\ and\ \citenamefont {Morello}}]{Laucht_arxiv_2015}%
  \BibitemOpen
  \bibfield  {author} {\bibinfo {author} {\bibfnamefont {A.}~\bibnamefont
  {Laucht}}, \bibinfo {author} {\bibfnamefont {J.~T.}\ \bibnamefont {Muhonen}},
  \bibinfo {author} {\bibfnamefont {F.~A.}\ \bibnamefont {Mohiyaddin}},
  \bibinfo {author} {\bibfnamefont {R.}~\bibnamefont {Kalra}}, \bibinfo
  {author} {\bibfnamefont {J.~P.}\ \bibnamefont {Dehollain}}, \bibinfo {author}
  {\bibfnamefont {S.}~\bibnamefont {Freer}}, \bibinfo {author} {\bibfnamefont
  {F.~E.}\ \bibnamefont {Hudson}}, \bibinfo {author} {\bibfnamefont
  {M.}~\bibnamefont {Veldhorst}}, \bibinfo {author} {\bibfnamefont
  {R.}~\bibnamefont {Rahman}}, \bibinfo {author} {\bibfnamefont
  {G.}~\bibnamefont {Klimeck}}, \bibinfo {author} {\bibfnamefont {K.~M.}\
  \bibnamefont {Itoh}}, \bibinfo {author} {\bibfnamefont {D.~N.}\ \bibnamefont
  {Jamieson}}, \bibinfo {author} {\bibfnamefont {J.~C.}\ \bibnamefont
  {McCallum}}, \bibinfo {author} {\bibfnamefont {A.~S.}\ \bibnamefont
  {Dzurak}}, \ and\ \bibinfo {author} {\bibfnamefont {A.}~\bibnamefont
  {Morello}},\ }\href@noop {} {\bibfield  {journal} {\bibinfo  {journal}
  {arXiv:1503.05985v1}\ } (\bibinfo {year} {2015})}\BibitemShut {NoStop}%
\bibitem [{\citenamefont {Wolfowicz}\ \emph {et~al.}(2013)\citenamefont
  {Wolfowicz}, \citenamefont {Tyryshkin}, \citenamefont {George}, \citenamefont
  {Riemann}, \citenamefont {Abrosimov}, \citenamefont {Becker}, \citenamefont
  {Pohl}, \citenamefont {Thewalt}, \citenamefont {Lyon},\ and\ \citenamefont
  {Morton}}]{Wolfowicz_NatureN_2013}%
  \BibitemOpen
  \bibfield  {author} {\bibinfo {author} {\bibfnamefont {G.}~\bibnamefont
  {Wolfowicz}}, \bibinfo {author} {\bibfnamefont {A.~M.}\ \bibnamefont
  {Tyryshkin}}, \bibinfo {author} {\bibfnamefont {R.~E.}\ \bibnamefont
  {George}}, \bibinfo {author} {\bibfnamefont {H.}~\bibnamefont {Riemann}},
  \bibinfo {author} {\bibfnamefont {N.~V.}\ \bibnamefont {Abrosimov}}, \bibinfo
  {author} {\bibfnamefont {P.}~\bibnamefont {Becker}}, \bibinfo {author}
  {\bibfnamefont {H.-J.}\ \bibnamefont {Pohl}}, \bibinfo {author}
  {\bibfnamefont {M.~L.~W.}\ \bibnamefont {Thewalt}}, \bibinfo {author}
  {\bibfnamefont {S.~A.}\ \bibnamefont {Lyon}}, \ and\ \bibinfo {author}
  {\bibfnamefont {J.~J.~L.}\ \bibnamefont {Morton}},\ }\href@noop {} {\bibfield
   {journal} {\bibinfo  {journal} {Nature Nanotechnolgy}\ }\textbf {\bibinfo
  {volume} {8}},\ \bibinfo {pages} {561} (\bibinfo {year} {2013})}\BibitemShut
  {NoStop}%
\bibitem [{\citenamefont {Boykin}\ \emph {et~al.}(2004)\citenamefont {Boykin},
  \citenamefont {Klimeck},\ and\ \citenamefont {Oyafuso}}]{Boykin_PRB_2004}%
  \BibitemOpen
  \bibfield  {author} {\bibinfo {author} {\bibfnamefont {T.~B.}\ \bibnamefont
  {Boykin}}, \bibinfo {author} {\bibfnamefont {G.}~\bibnamefont {Klimeck}}, \
  and\ \bibinfo {author} {\bibfnamefont {F.}~\bibnamefont {Oyafuso}},\
  }\href@noop {} {\bibfield  {journal} {\bibinfo  {journal} {Phys. Rev. B}\
  }\textbf {\bibinfo {volume} {69}},\ \bibinfo {pages} {115201} (\bibinfo
  {year} {2004})}\BibitemShut {NoStop}%
\bibitem [{Not()}]{Note_1}%
  \BibitemOpen
  \href@noop {} {}\bibinfo {note} {We have used the parameters given by Nara
  and Morita for P and As donors, and the parameters given by Pantelides and
  Sah for Sb and Bi donors, based on the best match with the experimental
  values for the binding energy spectra and $\eta_2$.}\BibitemShut {Stop}%
\bibitem [{\citenamefont {Ramdas}\ and\ \citenamefont
  {Rodriguez}(1981)}]{Ramdas_RPP_1981}%
  \BibitemOpen
  \bibfield  {author} {\bibinfo {author} {\bibfnamefont {A.~K.}\ \bibnamefont
  {Ramdas}}\ and\ \bibinfo {author} {\bibfnamefont {S.}~\bibnamefont
  {Rodriguez}},\ }\href@noop {} {\bibfield  {journal} {\bibinfo  {journal}
  {Rep. Prog. Phys.}\ }\textbf {\bibinfo {volume} {44}},\ \bibinfo {pages}
  {1297} (\bibinfo {year} {1981})}\BibitemShut {NoStop}%
\bibitem [{\citenamefont {Kohn}\ and\ \citenamefont
  {Luttinger}(1955)}]{Kohn_PR_1955}%
  \BibitemOpen
  \bibfield  {author} {\bibinfo {author} {\bibfnamefont {W.}~\bibnamefont
  {Kohn}}\ and\ \bibinfo {author} {\bibfnamefont {J.~M.}\ \bibnamefont
  {Luttinger}},\ }\href@noop {} {\bibfield  {journal} {\bibinfo  {journal}
  {Phys. Rev.}\ }\textbf {\bibinfo {volume} {98}},\ \bibinfo {pages} {915}
  (\bibinfo {year} {1955})}\BibitemShut {NoStop}%
\bibitem [{\citenamefont {Boykin}\ \emph {et~al.}(2002)\citenamefont {Boykin},
  \citenamefont {Klimeck}, \citenamefont {Bowen},\ and\ \citenamefont
  {Oyafuso}}]{Boykin_PRB_2002}%
  \BibitemOpen
  \bibfield  {author} {\bibinfo {author} {\bibfnamefont {T.~B.}\ \bibnamefont
  {Boykin}}, \bibinfo {author} {\bibfnamefont {G.}~\bibnamefont {Klimeck}},
  \bibinfo {author} {\bibfnamefont {R.~C.}\ \bibnamefont {Bowen}}, \ and\
  \bibinfo {author} {\bibfnamefont {F.}~\bibnamefont {Oyafuso}},\ }\href@noop
  {} {\bibfield  {journal} {\bibinfo  {journal} {Phys. Rev. B}\ }\textbf
  {\bibinfo {volume} {66}},\ \bibinfo {pages} {125207} (\bibinfo {year}
  {2002})}\BibitemShut {NoStop}%
\bibitem [{\citenamefont {McCamey}\ \emph {et~al.}(2010)\citenamefont
  {McCamey}, \citenamefont {Tol}, \citenamefont {Morley},\ and\ \citenamefont
  {Boehme}}]{McCamey_Science_2010}%
  \BibitemOpen
  \bibfield  {author} {\bibinfo {author} {\bibfnamefont {D.}~\bibnamefont
  {McCamey}}, \bibinfo {author} {\bibfnamefont {J.~V.}\ \bibnamefont {Tol}},
  \bibinfo {author} {\bibfnamefont {G.~W.}\ \bibnamefont {Morley}}, \ and\
  \bibinfo {author} {\bibfnamefont {C.}~\bibnamefont {Boehme}},\ }\href@noop {}
  {\bibfield  {journal} {\bibinfo  {journal} {Science}\ }\textbf {\bibinfo
  {volume} {330}},\ \bibinfo {pages} {1652} (\bibinfo {year}
  {2010})}\BibitemShut {NoStop}%
\bibitem [{\citenamefont {Ahmed}\ \emph {et~al.}(2009)\citenamefont {Ahmed}
  \emph {et~al.}}]{Ahmed_Enc_2009}%
  \BibitemOpen
  \bibfield  {author} {\bibinfo {author} {\bibfnamefont {S.}~\bibnamefont
  {Ahmed}} \emph {et~al.},\ }\href@noop {} {\bibfield  {journal} {\bibinfo
  {journal} {Springer Encyclopedia of Complexity and Systems Science (Berlin:
  Springer)}\ ,\ \bibinfo {pages} {p5745}} (\bibinfo {year}
  {2009})}\BibitemShut {NoStop}%
\bibitem [{\citenamefont {Fuhrer}\ \emph {et~al.}(2009)\citenamefont {Fuhrer},
  \citenamefont {Fuchsle}, \citenamefont {Reusch}, \citenamefont {Weber},\ and\
  \citenamefont {Simmons}}]{Fuhrer_Nanoletters_2009}%
  \BibitemOpen
  \bibfield  {author} {\bibinfo {author} {\bibfnamefont {A.}~\bibnamefont
  {Fuhrer}}, \bibinfo {author} {\bibfnamefont {M.}~\bibnamefont {Fuchsle}},
  \bibinfo {author} {\bibfnamefont {T.~C.~G.}\ \bibnamefont {Reusch}}, \bibinfo
  {author} {\bibfnamefont {B.}~\bibnamefont {Weber}}, \ and\ \bibinfo {author}
  {\bibfnamefont {M.~Y.}\ \bibnamefont {Simmons}},\ }\href@noop {} {\bibfield
  {journal} {\bibinfo  {journal} {Nanoletters}\ }\textbf {\bibinfo {volume}
  {9}},\ \bibinfo {pages} {707} (\bibinfo {year} {2009})}\BibitemShut {NoStop}%
\bibitem [{\citenamefont {Kettle}\ \emph {et~al.}(2006)\citenamefont {Kettle},
  \citenamefont {Goan},\ and\ \citenamefont {Smith}}]{Kettle_PRB_2006}%
  \BibitemOpen
  \bibfield  {author} {\bibinfo {author} {\bibfnamefont {L.}~\bibnamefont
  {Kettle}}, \bibinfo {author} {\bibfnamefont {H.-S.}\ \bibnamefont {Goan}}, \
  and\ \bibinfo {author} {\bibfnamefont {S.~C.}\ \bibnamefont {Smith}},\
  }\href@noop {} {\bibfield  {journal} {\bibinfo  {journal} {Phys. Rev. B}\
  }\textbf {\bibinfo {volume} {73}},\ \bibinfo {pages} {115205} (\bibinfo
  {year} {2006})}\BibitemShut {NoStop}%
\bibitem [{\citenamefont {George}\ \emph {et~al.}(2010)\citenamefont {George},
  \citenamefont {Witzel}, \citenamefont {Riemann}, \citenamefont {Abrosimov},
  \citenamefont {Notzel}, \citenamefont {Thewalt},\ and\ \citenamefont
  {Morton}}]{George_PRL_2010}%
  \BibitemOpen
  \bibfield  {author} {\bibinfo {author} {\bibfnamefont {R.}~\bibnamefont
  {George}}, \bibinfo {author} {\bibfnamefont {W.}~\bibnamefont {Witzel}},
  \bibinfo {author} {\bibfnamefont {H.}~\bibnamefont {Riemann}}, \bibinfo
  {author} {\bibfnamefont {N.~V.}\ \bibnamefont {Abrosimov}}, \bibinfo {author}
  {\bibfnamefont {N.}~\bibnamefont {Notzel}}, \bibinfo {author} {\bibfnamefont
  {M.~L.~W.}\ \bibnamefont {Thewalt}}, \ and\ \bibinfo {author} {\bibfnamefont
  {J.~J.~L.}\ \bibnamefont {Morton}},\ }\href@noop {} {\bibfield  {journal}
  {\bibinfo  {journal} {Phys. Rev. Lett.}\ }\textbf {\bibinfo {volume} {105}},\
  \bibinfo {pages} {067601} (\bibinfo {year} {2010})}\BibitemShut {NoStop}%
\end{thebibliography}
%

\end{document}